# Trust in AI and Implications for the AEC Research: A Literature Analysis

Newsha Emaminejad[1], Alexa Maria North[1], and Reza Akhavian, Ph.D., M.ASCE[1]

[1]Department of Civil, Construction, and Environmental Engineering, San Diego State University, 5500 Campanile Dr., San Diego, CA 92182; e-mails: {nemaminejad859; anorth3467; rakhavian@sdsu.edu}

## ABSTRACT

Engendering trust in technically acceptable and psychologically embraceable systems requires domain-specific research to capture unique characteristics of the field of application. The architecture, engineering, and construction (AEC) research community has been recently harnessing advanced solutions offered by artificial intelligence (AI) to improve project workflows. Despite the unique characteristics of work, workers, and workplaces in the AEC industry, the concept of trust in AI has received very little attention in the literature. This paper presents a comprehensive analysis of the academic literature in two main areas of "trust in AI" and "AI in the AEC", to explore the interplay between AEC projects' unique aspects and the sociotechnical concepts that lead to trust in AI. A total of 490 peer-reviewed scholarly articles are analyzed in this study. The main constituents of human trust in AI are identified from the literature and are characterized within the AEC project types, processes, and technologies.

## INTRODUCTION

Artificial Intelligence (AI) has recently gained tremendous traction in the architecture, engineering, and construction (AEC) industry. AI applications in architectural design (Darko et al. 2020), site logistic planning (Braun and Borrmann 2019), safety management (Baker et al. 2020), progress monitoring and productivity improvement (Sacks et al. 2020), and building operations and maintenance (López et al. 2013) have been studied extensively by researchers and (to a lesser extent) implemented by practitioners. There is a consensus among researchers that the technology adoption rate in the AEC industry is stagnant (Czarnowski et al. 2018). This is despite the fact that the industry faces grand challenges such as poor safety and productivity records that can be addressed using AI-enabled solutions similar to other industries (Baker et al. 2020; Delgado et al. 2019). In the information systems and implementation science, the absence of trust is a known hindrance to adopting new technologies (Danks 2019). For a technology such as AI with opaque back-end processes, and in the context of AEC which traditionally lags behind advanced technology, this is a more severe problem (Pan and Zhang 2021). AI algorithms are generally not easy to explain in layman terms and the processes between the input and the output are not sufficiently transparent to an ordinary end-user (Arrieta et al. 2020). In such a situation and in the absence of proven performance, a construction project team whose work has to be delivered under a predetermined time and budget limitations tends to refuse experimentations and continues to use methods that are trusted traditionally. This lack of confidence or "trust" to introduce new workflows can be traced back to both technical and psychological factors. Trustworthy AI, as a relatively new research paradigm, seeks to investigate mechanisms that enable building trust between AI-enabled systems and end-users, thus enhancing adoption levels (Siau and Wang 2018). This paper presents the results of a thorough investigation of the literature to develop a foundation for exploring factors that enhance the trust in, and thus the adoption of AI in the AEC industry.

## METHODOLOGY OF THE LITERATURE ANALYSIS

In this study, first a keyword search was conducted to find articles published from 1985 to 2021 on Google Scholar and Scopus search engines. The search keywords included a combination of "Trust in AI", "Trustworthy AI", "Trust in robots", "Ethical AI", "Transparent and explainable AI", "Reliable and safe AI", "Artificial Intelligence applications in construction management", "AI applications in construction management", "Robotics in construction management", and "Construction Automation". Next, manual filters were implemented to ensure that the papers to be analyzed are peer-reviewed and written in the English language to enable further screening if needed, and discuss directly related topics. Through these filtered search, a total of 490 articles were identified. Out of this total number, 210 articles are focused only on the applications of AI in the AEC industry, while the remaining discuss acceptance and trust in AI with no tie to the AEC concepts. Paper keywords were analyzed using VOSviewer software to establish co-occurrence maps in which the studies identified by their main focus are shown by circles linked together using lines with varying widths and lengths (Van Eck and Waltman 2013).

## TRUSTWORTHY AI

Trustworthy AI is an interdisciplinary field of research, involving disciplines such as computer science, human-computer interaction, human factors, robotics, engineering, management information systems, and psychology. In the past, the importance of the relationship between the end-user and AI-based processes was often overshadowed by pure technical advancements in the field (Hatami et al. 2019). More recently, however, experts have identified trust as an important element that determines adoption levels. Therefore, trustworthy AI research has evolved as a human-centered interdisciplinary field in which the needs, perceptions, and behaviors of human users are taken into account in the system's design (Canal et al. 2020).

Trust can be defined as a set of specific beliefs dealing with competence, integrity, predictability, and the willingness of someone to depend on another in a risky situation (Gefen et al. 2003). The concept of trust in AI and its unique aspects that are different from trust in other technologies have been extensively studied in the literature (Gillath et al. 2021; Li et al. 2008; Toreini et al. 2020). Theoretical studies indicate that factors that can affect the level of trust in technology systems can be categorized as those related to the user (e.g., expertise, attitudes towards robots), the hardware or machine (e.g., reliability, anthropomorphism), and the environment (e.g., characteristics of the team and task) (Lewis et al. 2018).

To create an overview of the parameters that influence trust in AI, the bibliometric data of the reviewed publications were fed into VOSviewer to generate maps of keywords and thematic co-occurrences. Figure 1 shows a network of the keywords where the size of the nodes is proportionate with the frequency of the keyword occurrence, and the distance between two nodes is inversely proportionate with the strength of the relation between the keywords in the literature analyzed. Additionally, the nodes and links are color-coded to reflect the publication age. Articles published before 2010 (limited in numbers relative to those published after 2010) were not included for more distinguishing color-coding). Major parameters identified in Figure 1 are explained below. In most cases, more than one term is used to describe a parameter to contextualize the concept, and/or to bundle closely related parameters that frequently co-appear in the literature. It is worth mentioning that these parameters are not mutually exclusive nor collectively exhaustive. In some cases they have semantic overlaps and not all of them are required to build trust.

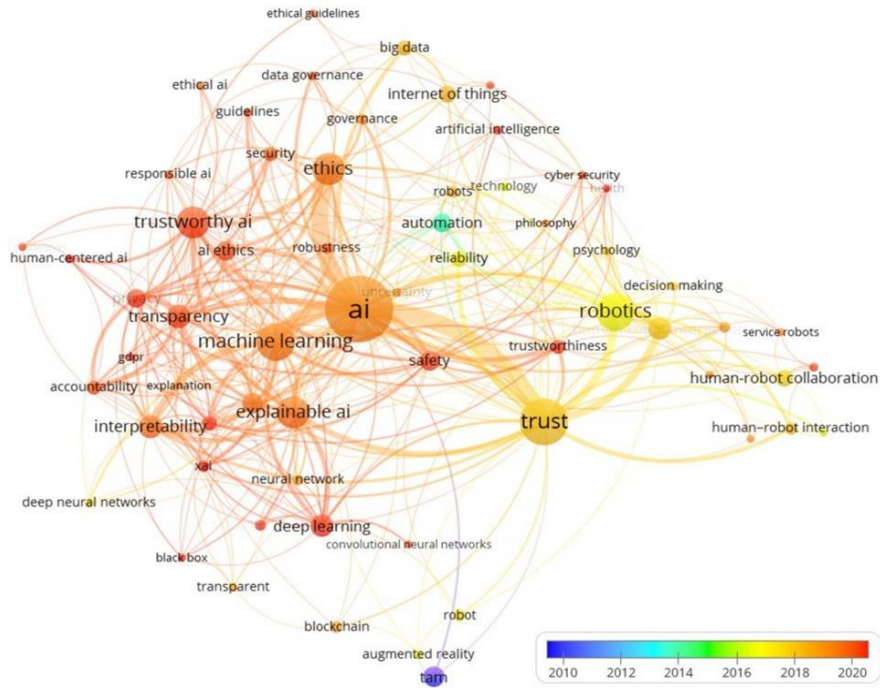

**Figure 1. Keyword co-occurrences network.**

**Transparency and Explainability (T&E).** In AI applications, transparency is often related to the concept of interpretability where the operations of a system can be understood by a human through introspection or explanation (Arrieta et al. 2020). Transparent and explainable AI systems are designed and implemented to be able to translate their operations into intelligible outputs that include information about how, when, and where they are used (Gillath et al. 2021; Toreini et al. 2020). The classic technology acceptance model (TAM) information systems theory (one of the oldest keywords in Figure 1) recognizes perception as a key element of technology adoption.

**Privacy and Security (P&S).** Humans' trust in technology is highly influenced by the levels of privacy involved in technology implementation (Li and Zhang 2017). Especially when sensitive (e.g., human health) data are in use, individuals' privacy and the risks of using the data both for the development of and the decisions made by the AI systems should be managed properly to enable trustworthiness (Yampolskiy 2018). To be trusted, AI systems must also be secure against being compromised by unauthorized agents (Siau and Wang 2018).

**Safety and Reliability (S&R).** AI systems that may pose risks of physical injury to users cannot be trusted (Tixier et al. 2016). This, of course, is a more important issue in embodied intelligence (e.g., intelligent robots), but even software and distributed computer networks will be distrusted should they manifest signs of safety and health threats (Baker et al. 2020). Reliability for trust is associated with the capacity of the models to avoid malfunctions; the vulnerabilities of AI models have to be identified and technical, or behavioral, solutions have to be implemented to ensure that autonomous systems will not be manipulated by an adversary (Ryan 2020).

**Ethics and Fairness (E&F).** Bias in training AI models, the ethical implications of developing biased intelligent systems, and the consequences of trusting and adopting them have been discussed extensively in the literature (Chakraborty et al. 2020; Siau and Wang 2018). Research has strongly advocated for diversity and inclusion to maximize fairness, minimize discrimination, and strengthen the basis of trust (Bartneck et al. 2021; Ryan and Stahl 2020; Toreini et al. 2020).

**Human-Centered Technology (HC).** A great deal of distrust in AI stems from a hypothesis that AI-powered technologies, such as intelligent robots, are developed to eliminate large segments of the workforce (Jarrahi 2018; Manzo et al. 2018). Many researchers conclude that to initiate trust between humans and AI, human-centered systems must be the cornerstone of research and development, where AI systems are designed to serve humankind, upskill workers, and promote human values (Dignum 2017; Lewis et al. 2018; Shneiderman 2020).

**Benevolence and Affect (BA).** It has been reported in behavioral science studies that the level of trust between humans and AI can be enhanced if they can bond socially and become "friends". Most of the studies focused on this concept fall in the broader category of human-AI or human-robot interaction (Pitardi and Marriott 2021; Toreini et al. 2020; Wang et al. 2016).

**TOWARD TRUSTWORTHY AI IN THE AEC RESEARCH AND PRACTICE**

The remarkable growth of the AI applications in the AEC domains has led to a number of review studies that highlight the status quo and the future potentials of the field. Most of these published studies focus on the value of implementing AI in a specific subfield, such as structural engineering (Salehi and Burgueño 2018), building information modeling (BIM) (Jianfeng et al. 2020), automated construction manufacturing (Hatami et al. 2019), and computer vision (Zhang et al. 2020). There are a limited number of studies that review general applications of AI in the AEC. Recent examples include Darko et al. (2020) and Pan and Zhang (2021) who reviewed and analyzed the use of AI in construction using a scientometric approach and identified the most commonly addressed topics, as well as future opportunities. However, to the best of the authors' knowledge, the topic of trust in AI with the AEC applications has never been studied before; neither as a literature analysis nor an independent research study. To present a comprehensive analysis of the applications of AI in the AEC literature through the lens of trust, this study identifies different categories within which the use of AI, and how it can be trusted, are substantially different. For example, engendering trust to leverage AI during the design phase calls for addressing "fairness" much more than "safety," while during the construction phase, "safety and reliability," and during the operations phase, "privacy and security," can have a more influential effect in building trust. Similarly, the project sector (e.g., building versus infrastructure), the objective within an AEC project (e.g., enhancing productivity versus safety), and the technology that is powered by AI (e.g., BIM versus robotics) are important factors that can determine the approach towards establishing trust. Table 1 shows this categorization proposed based on the frequency of these categories in the keywords of the AI in AEC literature analyzed.

**Table 1. Proposed AEC projects categories and subcategories for trustworthy AI studies.**

| Category | Subcategories |
|---|---|
| Project Phase | Pre-Construction, Construction, Post-Construction |
| Construction Type | Building Construction, Horizontal Construction |
| Application | Safety, Productivity, Sustainability, Scheduling |
| Technology | BIM, Robotics, Mobile Computing, Blockchain |

**RESULTS AND DISCUSSION**

To understand the growth of research on the topic of trustworthy AI (and the lack thereof within the AEC industry), the results of the conducted literature analysis are tabulated and visualized in this section. A temporal bibliographic analysis is provided in Figure 2 (a), where a histogram of publication dates of articles reviewed in this study is presented in two categories: those discussing trust in AI, and those focused on developing or adopting AI models for AEC applications. With 441 reviewed papers for both categories from 2005 to 2021 (out of the total 490 from 1985), rapid growth can be seen in this right-skewed histogram, as more than 92% of these papers have been published between 2016 and 2021. Furthermore, as the interest in leveraging AI in the AEC industry grows, the number of studies targeting trust in AI (in fields other than AEC) is also on the rise. This can indicate a major demand for this research in the future within the AEC community.

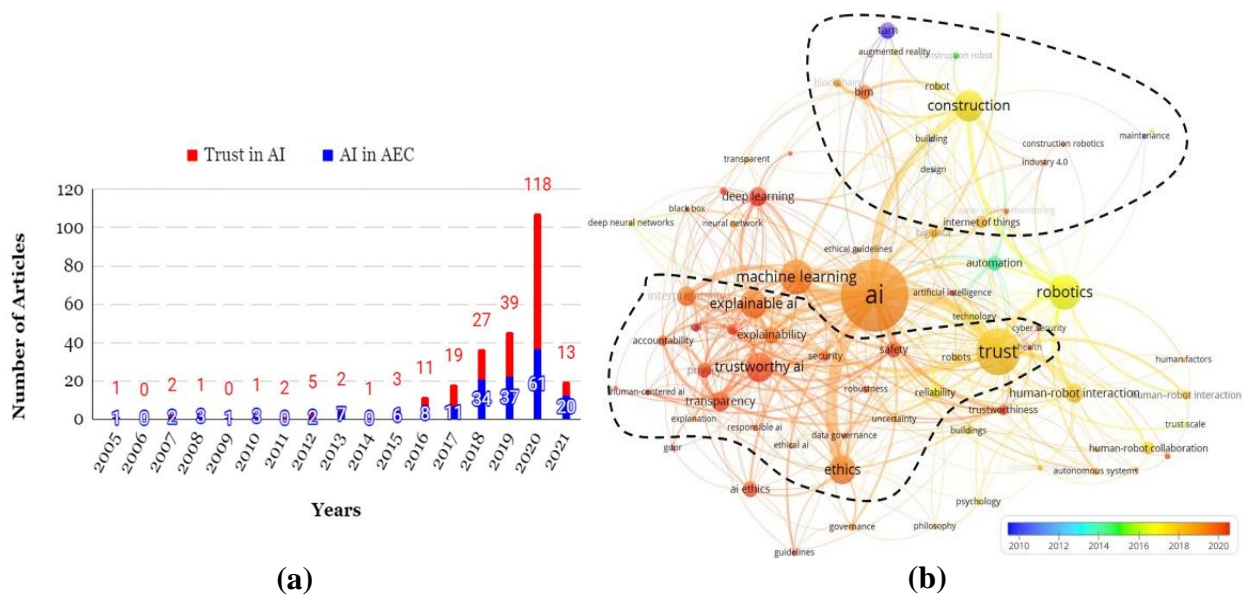

(a) (b)

**Figure 2 (a). A histogram of the analyzed papers' publication years; (b). A comprehensive network of the keywords with two distant clusters.**

Another informative visualization is a comprehensive view of the VOSviewer network, presented in Figure 2 (b). The top cluster indicating studies pertinent to AEC with major connections to topics such as AI, machine learning, deep learning, and robotics is distant and virtually disconnected from the trust keywords clustered below it. In addition, a comparison between Figure 1 and Figure 2 (b) reveals that the relatively older topic of TAM (the blue circle on top) is now appearing closer to the AEC topics, with direct links to topics such as BIM and augmented reality in the top cluster and AI topics. This indicates that TAM has received substantial attention in both the AEC and the trust in AI literature for technology adoption, and can play a major role in theoretical as well as empirical studies related to AI in the AEC.

Finally, a thematic investigation of the literature analyzed in this paper in terms of the trustworthy AI parameters identified in Figure 1, as well as the proposed categories of the AEC concepts in Table 1, can help better identify the existing and future potentials of this topic. The subcategories indicated in Table 2 are those identified explicitly in the papers analyzed. The applicable trust parameters, however, are determined through a thematic analysis performed by the research team members, based on the content of the papers classified in each category. It is worth mentioning that the sum of the number of papers in Table 2 is more than the total number of papers reviewed since some papers discuss more than one subcategory.

**Table 2. Thematic analysis of the reviewed papers with identified subcategories and applicable trust parameters.**

| Subcategory | # of Papers | Applicable Trust Parameters | | | | | |
|---|---|---|---|---|---|---|---|
| | | T&E | P&S | S&R | E&F | H&C | B&A |
| Pre-Construction | 19 | ✓ | ✓ | ✓ | ✓ | ✓ | |
| Construction | 131 | ✓ | ✓ | ✓ | ✓ | ✓ | ✓ |
| Post-Construction | 32 | ✓ | ✓ | ✓ | ✓ | | |
| Building Construction | 29 | ✓ | ✓ | ✓ | ✓ | ✓ | ✓ |
| Infrastructure Projects | 34 | ✓ | ✓ | ✓ | | ✓ | |
| Safety | 12 | ✓ | ✓ | ✓ | ✓ | ✓ | |
| Productivity | 6 | ✓ | ✓ | ✓ | ✓ | ✓ | ✓ |
| Sustainability | 15 | ✓ | ✓ | ✓ | ✓ | ✓ | ✓ |
| Scheduling | 8 | ✓ | ✓ | ✓ | | ✓ | |
| BIM | 12 | ✓ | ✓ | ✓ | | | |
| Robotics | 74 | ✓ | ✓ | ✓ | ✓ | ✓ | ✓ |
| Mobile Computing | 13 | ✓ | ✓ | ✓ | ✓ | ✓ | |
| Blockchain | 14 | ✓ | ✓ | | ✓ | | ✓ |

**CONCLUSION**

A detailed cross-referencing of the papers analyzed in this research through temporal study, keyword co-occurrence network creation, and thematic analysis indicates a substantial potential and need for exploring trust concepts in adopting AI in AEC applications. The interdisciplinary nature of the topic allows for observing the interplay of the subcategories and parameters identified in this paper from the lens of different disciplinary fields. For example, from the AEC perspective and within the project phase category, "Construction" appears to have the potential to engage all the trust parameters identified. A similar argument is valid for "Robotics" within the technology category. From a trust in AI perspective, transparency and explainability are identified as major trust dimensions across all AEC subcategories. Safety and reliability, and to a lesser extent privacy and security, as well as ethics and fairness, are trust concepts that can be widely applied to the use of AI in AEC research and practice.

      The presented study describes preliminary findings of a larger research endeavor to study trust development and calibration for AI in AEC applications. The study has a few limitations that can be addressed in future work. The thematic analysis presented in Table 2 is a subjective assessment of the research team. Further analysis can incorporate survey or interview results with experts in the field. Additionally, search keywords can be expanded to incorporate subtopics of AI that may have been listed instead of AI keywords in the original publications.

# REFERENCES

Arrieta, A. B., Díaz-Rodríguez, N., Del Ser, J., Bennetot, A., Tabik, S., Barbado, A., García, S., Gil-López, S., Molina, D., and Benjamins, R. (2020). "Explainable Artificial Intelligence (XAI): Concepts, taxonomies, opportunities and challenges toward responsible AI." *Information Fusion*, 58, 82-115.

Baker, H., Hallowell, M. R., and Tixier, A. J.-P. (2020). "AI-based prediction of independent construction safety outcomes from universal attributes." *AutCon*, 118, 103146.

Bartneck, C., Lütge, C., Wagner, A., and Welsh, S. (2021). *An introduction to ethics in robotics and AI*, Springer Nature.

Braun, A., and Borrmann, A. (2019). "Combining inverse photogrammetry and BIM for automated labeling of construction site images for machine learning." *AutCon*, 106, 102879.

Canal, G., Borgo, R., Coles, A., Drake, A., Huynh, D., Keller, P., Krivić, S., Luff, P., Mahesar, Q.-a., and Moreau, L. (2020). "Building Trust in Human-Machine Partnerships⋆." ELSEVIER ADVANCED TECHNOLOGY OXFORD FULFILLMENT CENTRE THE BOULEVARD ....

Chakraborty, J., Peng, K., and Menzies, T. "Making fair ML software using trustworthy explanation." *Proc., 2020 35th IEEE/ACM International Conference on Automated Software Engineering (ASE)*, IEEE, 1229-1233.

Czarnowski, J., Dąbrowski, A., Maciaś, M., Główka, J., and Wrona, J. (2018). "Technology gaps in human-machine interfaces for autonomous construction robots." *AutCon*, 94, 179-190.

Danks, D. "The value of trustworthy AI." *Proc., Proceedings of the 2019 AAAI/ACM Conference on AI, Ethics, and Society*, 521-522.

Darko, A., Chan, A. P., Adabre, M. A., Edwards, D. J., Hosseini, M. R., and Ameyaw, E. E. (2020). "Artificial intelligence in the AEC industry: Scientometric analysis and visualization of research activities." *AutCon*, 112, 103081.

Delgado, J. M. D., Oyedele, L., Ajayi, A., Akanbi, L., Akinade, O., Bilal, M., and Owolabi, H. (2019). "Robotics and automated systems in construction: Understanding industry-specific challenges for adoption." *Journal of Building Engineering*, 26, 100868.

Dignum, V. (2017). "Responsible artificial intelligence: designing AI for human values."

Gefen, D., Karahanna, E., and Straub, D. W. (2003). "Trust and TAM in online shopping: An integrated model." *MIS quarterly*, 51-90.

Gillath, O., Ai, T., Branicky, M. S., Keshmiri, S., Davison, R. B., and Spaulding, R. (2021). "Attachment and trust in artificial intelligence." *Computers in Human Behavior*, 115, 106607.

Hatami, M., Flood, I., Franz, B., and Zhang, X. (2019). "State-of-the-Art Review on the Applicability of AI Methods to Automated Construction Manufacturing." *Computing in Civil Engineering 2019: Data, Sensing, and Analytics*, 368-375.

Jarrahi, M. H. (2018). "Artificial intelligence and the future of work: Human-AI symbiosis in organizational decision making." *Business Horizons*, 61(4), 577-586.

Jianfeng, Z., Yechao, J., and Fang, L. "Construction of Intelligent Building Design System Based on BIM and AI." *Proc., 2020 5th International Conference on Smart Grid and Electrical Automation (ICSGEA)*, IEEE, 277-280.

Lewis, M., Sycara, K., and Walker, P. (2018). "The role of trust in human-robot interaction." *Foundations of trusted autonomy*, Springer, Cham, 135-159.

Li, X., Hess, T. J., and Valacich, J. S. (2008). "Why do we trust new technology? A study of initial trust formation with organizational information systems." *The Journal of Strategic Information Systems*, 17(1), 39-71.
Li, X., and Zhang, T. "An exploration on artificial intelligence application: From security, privacy and ethic perspective." *Proc., 2017 IEEE 2nd International Conference on Cloud Computing and Big Data Analysis (ICCCBDA)*, IEEE, 416-420.
López, J., Pérez, D., Paz, E., and Santana, A. (2013). "WatchBot: A building maintenance and surveillance system based on autonomous robots." *Robotics and Autonomous Systems*, 61(12), 1559-1571.
Manzo, J., Manzo, F., and Bruno, R. (2018). "The potential economic consequences of a highly automated construction industry." *What If Construction Becomes the Next Manufacturing*.
Pan, Y., and Zhang, L. (2021). "Roles of artificial intelligence in construction engineering and management: A critical review and future trends." *AutCon*, 122, 103517.
Pitardi, V., and Marriott, H. R. (2021). "Alexa, she's not human but… Unveiling the drivers of consumers' trust in voice-based artificial intelligence." *Psychology & Marketing*, 38(4), 626-642.
Ryan, M. (2020). "In AI We Trust: Ethics, Artificial Intelligence, and Reliability." *Science and Engineering Ethics*, 26(5), 2749-2767.
Ryan, M., and Stahl, B. C. (2020). "Artificial intelligence ethics guidelines for developers and users: clarifying their content and normative implications." *Journal of Information, Communication and Ethics in Society*.
Sacks, R., Girolami, M., and Brilakis, I. (2020). "Building information modelling, artificial intelligence and construction tech." *Developments in the Built Environment*, 4, 100011.
Salehi, H., and Burgueño, R. (2018). "Emerging artificial intelligence methods in structural engineering." *Engineering structures*, 171, 170-189.
Shneiderman, B. (2020). "Human-centered artificial intelligence: Reliable, safe & trustworthy." *International Journal of Human–Computer Interaction*, 36(6), 495-504.
Siau, K., and Wang, W. (2018). "Building trust in artificial intelligence, machine learning, and robotics." *Cutter Business Technology Journal*, 31(2), 47-53.
Tixier, A. J.-P., Hallowell, M. R., Rajagopalan, B., and Bowman, D. (2016). "Application of machine learning to construction injury prediction." *AutCon*, 69, 102-114.
Toreini, E., Aitken, M., Coopamootoo, K., Elliott, K., Zelaya, C. G., and van Moorsel, A. "The relationship between trust in AI and trustworthy machine learning technologies." *Proc., Proceedings of the 2020 Conference on Fairness, Accountability, and Transparency*, 272-283.
Toreini, E., Aitken, M., Coopamootoo, K. P., Elliott, K., Zelaya, V. G., Missier, P., Ng, M., and van Moorsel, A. (2020). "Technologies for Trustworthy Machine Learning: A Survey in a Socio-Technical Context." *arXiv preprint arXiv:2007.08911*.
Van Eck, N. J., and Waltman, L. (2013). "VOSviewer manual." *Leiden: Univeristeit Leiden*, 1(1), 1-53.
Wang, N., Pynadath, D. V., and Hill, S. G. "The impact of pomdp-generated explanations on trust and performance in human-robot teams." *Proc., Proceedings of the 2016 international conference on autonomous agents & multiagent systems*, 997-1005.
Yampolskiy, R. V. (2018). *Artificial intelligence safety and security*, CRC Press.
Zhang, Y., Liu, H., Kang, S.-C., and Al-Hussein, M. (2020). "Virtual reality applications for the built environment: Research trends and opportunities." *AutCon*, 118, 103311.